\makeatletter
\@ifundefined{@parse@version@dash}{%
\def\@parse@version#1{\@parse@version@0#1}
\def\@parse@version@#1/#2/#3#4#5\@nil{%
\@parse@version@dash#1-#2-#3#4\@nil}
\def\@parse@version@dash#1-#2-#3#4#5\@nil{%
  \if\relax#2\relax\else#1\fi#2#3#4 }
}{}
\makeatother

\documentclass[aps, prb, twocolumn, reprint, superscriptaddress, longbibliography]{revtex4-2}
\usepackage{bm}
\usepackage{color}
\usepackage{graphicx}
\usepackage{adjustbox}

\date{\today}

\begin{document}


\title{Sublattice spin reversal and field induced Fe\textsuperscript{3+} spin-canting across the magnetic compensation temperature in Y\textsubscript{1.5}Gd\textsubscript{1.5}Fe\textsubscript{5}O\textsubscript{12} rare-earth iron garnet}

\author{Manik Kuila}
\affiliation{UGC-DAE Consortium for Scientific Research, University Campus, Khandwa Road, Indore 452001, India.}

\author{Jose Mardegan}
\affiliation{Deutsches Elektronen-Synchrotron DESY, Notkestraße 85, 22607 Hamburg, Germany.}

\author{Akhil Tayal}
\affiliation{Deutsches Elektronen-Synchrotron DESY, Notkestraße 85, 22607 Hamburg, Germany.}

\author{Sonia Francoual}
\affiliation{Deutsches Elektronen-Synchrotron DESY, Notkestraße 85, 22607 Hamburg, Germany.}

\author{V.Raghavendra Reddy}
\email{varimalla@yahoo.com; vrreddy@csr.res.in}
\affiliation{UGC-DAE Consortium for Scientific Research,  University Campus, Khandwa Road, Indore 452001, India.}

\begin{abstract}

In the present work Fe\textsuperscript{3+} sublattice spin reversal and Fe\textsuperscript{3+} spin-canting across the magnetic compensation temperature (T\textsubscript{Comp}) are demonstrated in polycrystalline Y\textsubscript{1.5}Gd\textsubscript{1.5}Fe\textsubscript{5}O\textsubscript{12} (YGdIG) by means of in-field $^{57}Fe$ M$\ddot{o}$ssbauer spectroscopy measurements. Corroborating in-field $^{57}Fe$ M$\ddot{o}$ssbauer measurements, both Fe\textsuperscript{3+} \& Gd\textsuperscript{3+} sublattice spin reversal has also been manifested with x-ray magnetic circular dichroism (XMCD) measurement in hard x-ray region. Moreover from in-field $^{57}Fe$ M$\ddot{o}$ssbauer measurements, estimation and analysis of effective internal hyperfine field (H\textsubscript{eff}), relative intensity of absorption lines in a sextet elucidated unambiguously the signatures of Fe\textsuperscript{3+} spin reversal, their continuous transition and field induced spin-canting of Fe\textsuperscript{3+} sublattices across T\textsubscript{Comp}. Further, Fe K- (Gd L\textsubscript{3}-) edge XMCD signal is observed to consist of additional spectral features, those are identified from Gd\textsuperscript{3+} (Fe\textsuperscript{3+}) magnetic ordering, enabling us the extraction of both the sublattices (Fe\textsuperscript{3+} \& Gd\textsuperscript{3+}) information from a single edge analysis. The evolution of the magnetic moments as a function of temperature for both magnetic sublattices extracted either at the Fe K- or Gd L3-edge agree quite well with values that are extracted from bulk magnetization data of YGdIG and YIG (Y\textsubscript{3}Fe\textsubscript{5}O\textsubscript{12}). These measurements pave new avenues to investigate how the magnetic behavior of such complex system acts across the compensation point.  

\end{abstract}


\maketitle \section{Introduction}

Rare-earth iron garnets (RIG), R\textsubscript{3}Fe\textsubscript{5}O\textsubscript{12}, where R is rare-earth (Y, \& La-Lu) have become an important class of ferrimagnetic oxide materials finding a significant role in many microwave, bubble memories and magneto-optical device applications \cite{microwavedevice, bubblememory1,bubblememory2, MOdevice}. Remarkable intriguing magnetic properties and their chemical stability with a large variety of elemental substitutions are one of the prime reasons for exploring these materials by various groups since their discovery \cite{Neelmagnetism_science, GarnetReview1, sublatticeMag1968, experimentalprofXMCD, finiteTempMagPro2020, TuningTcompCurie2019, DW_ElectricPolarization}. Among these RIG systems, yttrium iron garnet (Y\textsubscript{3}Fe\textsubscript{5}O\textsubscript{12}, YIG) received a renewed attention in recent years owing to its low damping, low optical absorption, magneto-optical switching, thermoelectric generation in spin Seebeck insulators and other spintronics, magneonic based application \cite{SpinDiffusionLength,MORecordingYIG,SpinSeebeckInsulator,Magnonvalve}. Doping of magnetic light rare-earths (e.g., Ce, Nd and Gd) or even heavier metals, such as Bi atoms, on the yttrium sites yield a significantly enhanced visible to near-infrared Faraday$/$Kerr rotation and magneto-optical (MO) figure of merit, without losing their magnetic insulator character \cite{enhancedMO_PRAppl,MOCeYIG_FOM,Manik_JAP,MOresponseGdIG,BIGMOresponse}. 

RIG consist of three different magnetic ions viz., two in-equivalent Fe\textsuperscript{3+}ions located at tetrahedral ($\textit{d}$-) and octahedral ($\textit{a}$-) oxygen polyhedron and the third one is the R\textsuperscript{3+} ion situated at dodecahedral ($\textit{c}$) oxygen polyhedron \cite{Neelmagnetism_science}. The $\textit{d}$- and $\textit{a}$- site Fe\textsuperscript{3+} are always coupled anti-parallel and the resultant moment of Fe\textsuperscript{3+} is also coupled anti-parallel with the R\textsuperscript{3+}. These two magnetic sublattices exhibit quite a different temperature dependance and as a result there exists a point in temperature at which the resultant Fe\textsuperscript{3+} magnetic moments are equal and opposite to the R\textsuperscript{3+} magnetic moments resulting in zero total magnetization \cite{GarnetReview1}, known as magnetic compensation temperature (T\textsubscript{Comp}). Therefore, a thorough understanding of the macroscopic magnetic properties in RIG compounds is achieved from the knowledge of the different sublattices magnetization \cite{Cornellius_PRB2012,Cornellius_PRL1,Neelferri_largeField, SoftXMCD, SoftXMCD_HoIG,SublatticeMOresponse,untanglingCeYIG}. 

\begin{figure}[t]
\centering
\includegraphics[height = 5 cm, width=8 cm, keepaspectratio]{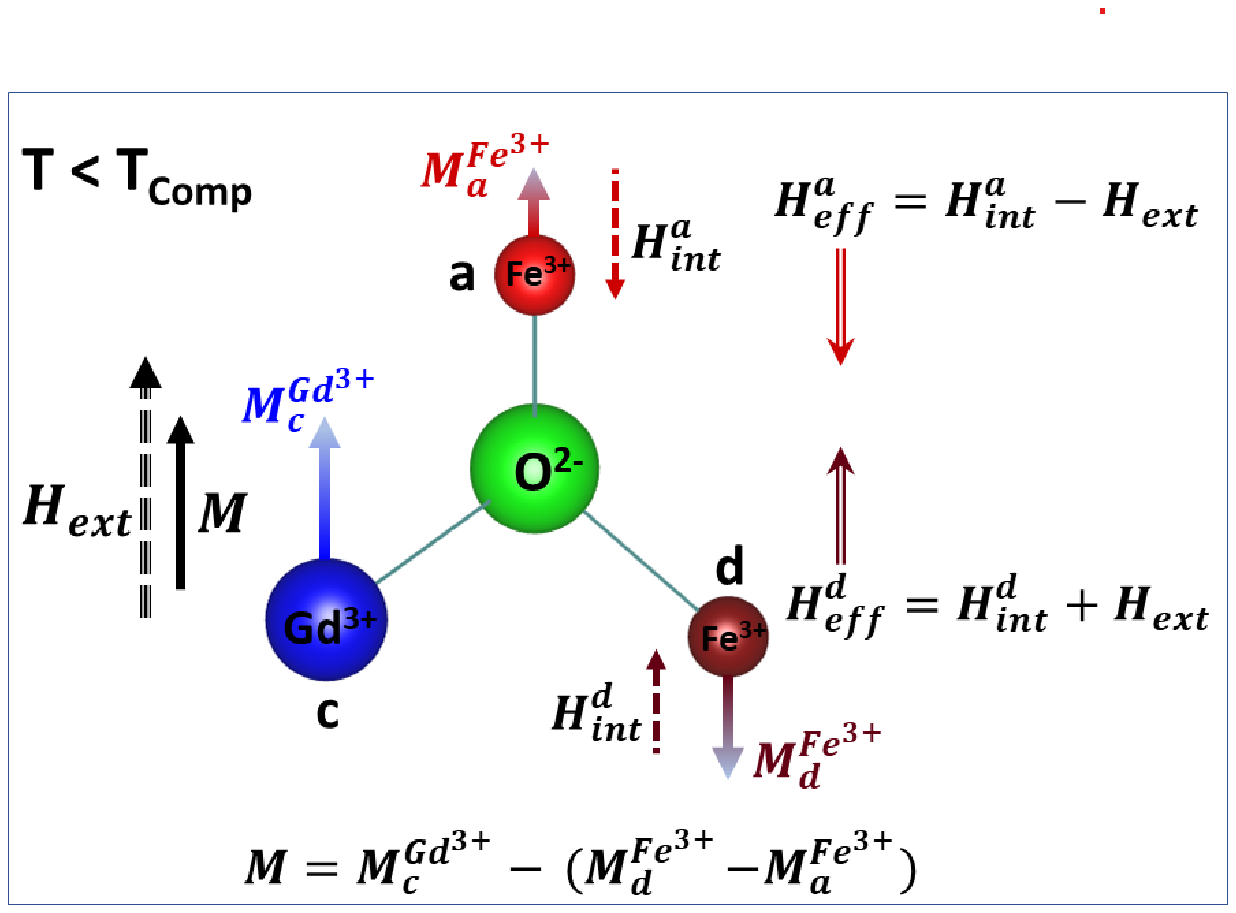}
\includegraphics[height = 5 cm, width=6.9 cm, keepaspectratio]{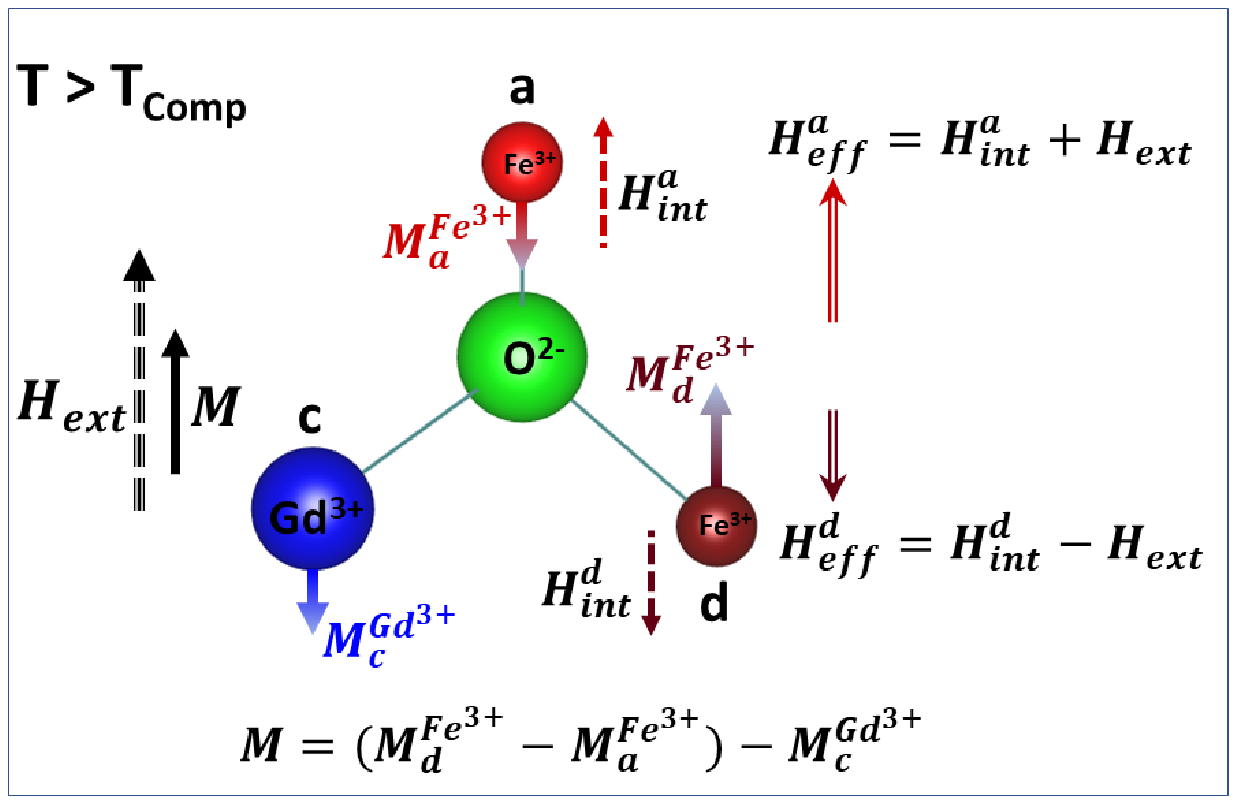}
\caption{Schematic diagram showing the relation between internal (H\textsubscript{int}), effective (H\textsubscript{eff}) hyperfine and externally applied (H\textsubscript{ext}) fields below and above magnetic compensation (T\textsubscript{Comp}) temperature. M's represent the magnetic moments and the corresponding arrows are direction of the moments of the respective sublattice.}
\label{fig:cartoon}
\end{figure}

Usually at low temperatures i.e., T$<$T\textsubscript{Comp} the R\textsuperscript{3+} moments dominate the macroscopic magnetic properties and align along the externally applied magnetic field (H\textsubscript{ext}), whereas for intermediate temperature, T\textsubscript{Comp}$<$T$<$T\textsubscript{C} (Curie temperature) it is the resultant Fe\textsuperscript{3+} moments which will dominate and align along the H\textsubscript{ext}. The net magnetization (M) always align along the H\textsubscript{ext}. Therefore, the magnetic sub-lattices rearrange accordingly whether the temperature of the system is below or above the T\textsubscript{Comp} as shown schematically in Figure.~\ref{fig:cartoon}. Now, if the R-atom is non-magnetic (like as in YIG systems) or for temperatures above the T\textsubscript{Comp}, the dominant Fe\textsuperscript{3+} at $\textit{d}$- sites decide the resultant magnetization direction, whereas for temperatures below T\textsubscript{Comp}, the $\textit{d}$- site Fe\textsuperscript{3+} moments will be in opposite direction to H\textsubscript{ext}. However, the competition between H\textsubscript{ext} that always tend to align all the moments parallel to it and the strong anti-ferromagnetic super-exchange interaction between the sublattices result field induced spin-canted phase (between magnetic R\textsuperscript{3+} and resultant Fe\textsuperscript{3+}) close to T\textsubscript{Comp}, which is shown to be a second-order phase transition in literature \cite{Neelferri_largeField,Cornellius_PRB2012,Cornellius_PRL1}.

In this context, the in-field $^{57}Fe$ M$\ddot{o}$ssbauer spectroscopy is an ideal method to track the evolution of Fe\textsuperscript{3+} sublattice magnetization, demonstrate their inversion across T\textsubscript{Comp} and also probe the spin-canting unambiguously by analyzing the variation of effective internal field (H\textsubscript{eff}), which is the vector sum of internal hyperfine field (H\textsubscript{int}), H\textsubscript{ext} \& the relative line intensities in a given six-line pattern \cite{LTHMMoss1, LTHMMoss2}. However, there seems to be no systematic in-field M$\ddot{o}$ssbauer study across T\textsubscript{Comp} in RIG systems in literature. Most of the M$\ddot{o}$ssbauer literature focused on the estimation of cation distribution in RIG systems, except the work of Stadnik et al., and Seidel et al \cite{HFI_Seidel, InfieldMoss}. Seidel et al., reported that there is no spin-canting in Gd\textsubscript{3}Fe\textsubscript{5}O\textsubscript{12} across T\textsubscript{Comp} with zero-field M$\ddot{o}$ssbauer measurements \cite{HFI_Seidel} and Stadnik et al., reported canting of the Fe\textsuperscript{3+} spins in Sc doped Eu\textsubscript{3}Fe\textsubscript{5}O\textsubscript{12} with in-field M$\ddot{o}$ssbauer measurements \cite{InfieldMoss}.


The in-field $^{57}Fe$ M$\ddot{o}$ssbauer measurements reported in this paper, are further complimented by x-ray magnetic circular dichroism (XMCD) measurements. In RIG systems, most of the XMCD investigation is focused on the transition metal (TM) L\textsubscript{2,3} (2p$\rightarrow$3d) and R- M\textsubscript{4,5} (3d$\rightarrow$4f) absorption edges, since in both cases the magnetism is directly probed via electric-dipole transitions \cite{SoftXMCD, SoftXMCD_HoIG}. Recently studies have shown that due to significant improvement in the 3rd generation synchrotron, the XMCD method is also possible to be carried out with much higher signal to noise ratio even at hard x-rays regime i.e., across the Fe-K and R L-edges \cite{HXMCD_Review}. Further it is shown in the previous literature that Fe K-edge (R L-edge) XMCD signal is strongly influenced by the R-atom (Fe-atom) magnetic ordering in rare-earth iron compounds such as RFe\textsubscript{2}, NdFeB and RIGs. Therefore, by performing XMCD measurements in hard x-ray region across a single edge, one would be able to get the element specific magnetic information from both Fe and R sublattices in these type of compounds \cite{Cornellius_PRB2012, Chaboy_PRB2007, Chaboy_PRB2008, Cornellius_PRL1, Cornellius_PRL2}. 

\begin{figure}[b]
  \centering
  \includegraphics[scale=1, width = 8 cm]{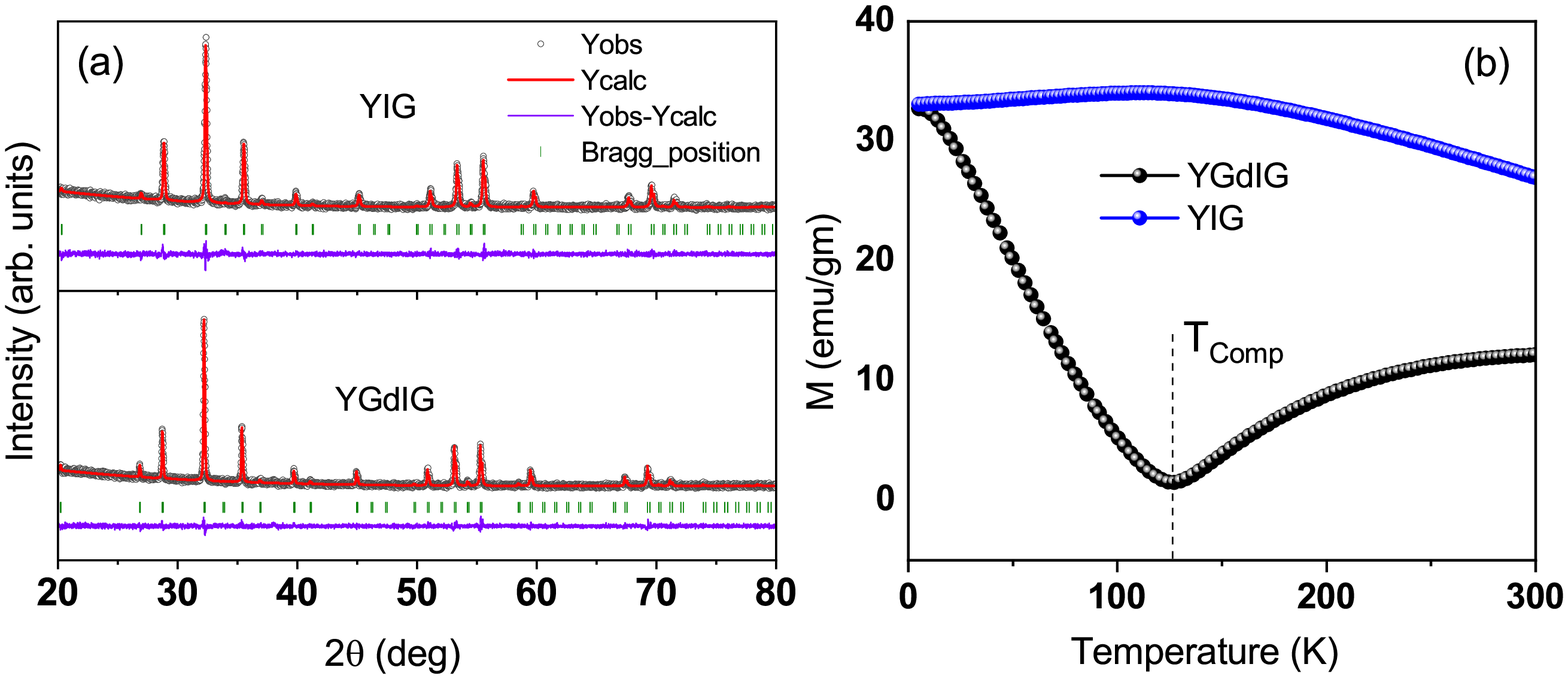}
  \caption{(a) X-ray diffraction data of Y\textsubscript{1.5}Gd\textsubscript{1.5}Fe\textsubscript{5}O\textsubscript{12} and Y\textsubscript{3}Fe\textsubscript{5}O\textsubscript{12}  (b) M-T data measured in field-cooled protocol with 500 Oe field. The magnetic compensation (T\textsubscript{Comp}) is clearly seen for YGdIG as indicated by black dash line.}
	\label{fig:XRD_Mag}
\end{figure}

The present work reports the temperature dependent in-field $^{57}Fe$ M$\ddot{o}$ssbauer spectroscopy and XMCD measurements on polycrystalline Y\textsubscript{1.5}Gd\textsubscript{1.5}Fe\textsubscript{5}O\textsubscript{12} (YGdIG) and Y\textsubscript{3}Fe\textsubscript{5}O\textsubscript{12} (YIG) samples with the aim of tracing sublattice spin across T\textsubscript{Comp}, looking for the possible spin-canting across T\textsubscript{Comp} and decomposition of XMCD data into Gd-like and Fe-like spectra from a single edge (i.e., Gd L\textsubscript{3}-edge or Fe K-edge) measurement.

\section{Experimental}

Polycrystalline Y\textsubscript{1.5}Gd\textsubscript{1.5}Fe\textsubscript{5}O\textsubscript{12} and Y\textsubscript{3}Fe\textsubscript{5}O\textsubscript{12} samples are prepared with conventional solid-state-reaction method starting with high purity ($\geq$99.9\%) oxide precursors. The structural characterization of the prepared samples is carried out with Brucker D8-Discover x-ray diffraction system equipped with LynxEye detector and Cu-K\textsubscript{$\alpha$} source. $^{57}Fe$ M$\ddot{o}$ssbauer measurements are carried out in transmission mode using a standard PC-based M$\ddot{o}$ssbauer spectrometer equipped with a WissEl velocity drive in constant acceleration mode. The velocity calibration of the spectrometer is done with natural iron absorber at room temperature. For the low temperature high magnetic field M$\ddot{o}$ssbauer measurements, the sample is placed inside a Janis superconducting magnet and an H\textsubscript{ext} is applied parallel to the $\gamma$-rays.  Bulk magnetization measurements are performed with vibrating sample magnetometer (VSM). X-ray absorption spectroscopy (XAS) measurements are carried out at beamline P09 at PETRA III (DESY) at low temperature at the Gd L\textsubscript{3} and Fe K-absorption edges in transmission geometry in which both incident and transmitted x-ray beams are recorded using silicon photo-diodes. The pellets, prepared from samples with suitable amount of boron nitrate add-mixer, are cooled down by an ARS cryostat with temperature range between 5 to 300 K. XAS$/$XMCD measurements were performed fast-switching the beam helicity between left and right circular polarization to improve the signal-to-noise ratio. The dichroic signal is revealed when the difference in the absorption spectra for left and right circularly polarized radiation is performed \cite{P09}. In order to align the domains and to correct for nonmagnetic artifacts in the XMCD data, an external magnetic field of approximately 1 T was applied parallel and antiparallel to the incident beam wave vector $\stackrel{\rightarrow}{k}$. 

\section{Results}

\subsection{Structural and magnetic characterization}
 
Figure.~\ref{fig:XRD_Mag}(a) shows the XRD patterns of the YIG and YGdIG samples and the data confirms the phase purity of the prepared samples. Further, the XRD data is analyzed with FullProf Rietveld refinement \cite{FullProf} program considering the $\textit{Ia}$-3$\textit{d}$ space-group for the estimation of lattice parameters and the obtained lattice parameters are 12.374(1) $\dot{A}$ and 12.424(1) $\dot{A}$ for YIG and YGdIG, respectively, which also match with the literature \cite{LatticeparameterRIG}. Figure.~\ref{fig:XRD_Mag}(b) shows the temperature dependent (M-T) magnetization data of the YIG and YGdIG samples. One can clearly see the magnetic compensation temperature (T\textsubscript{Comp}) for YGdIG at about 126 K and no such signature is seen for YIG as expected. Since, the bulk magnetization data of YIG is considered to be coming only from Fe\textsuperscript{3+} moments this gives a rough idea about the contribution of Fe\textsuperscript{3+} sublattice to the total magnetization in YGdIG.  So, one can get an estimated contribution of Gd\textsuperscript{3+} sublattice to the overall moment in YGdIG by suitably subtracting the bulk magnetization data of YGdIG from YIG data. This method is employed to cross-check the reliability of the XMCD data analysis results as discussed in the later sections. 

\begin{figure}[b]
\centering
\includegraphics[width=8 cm]{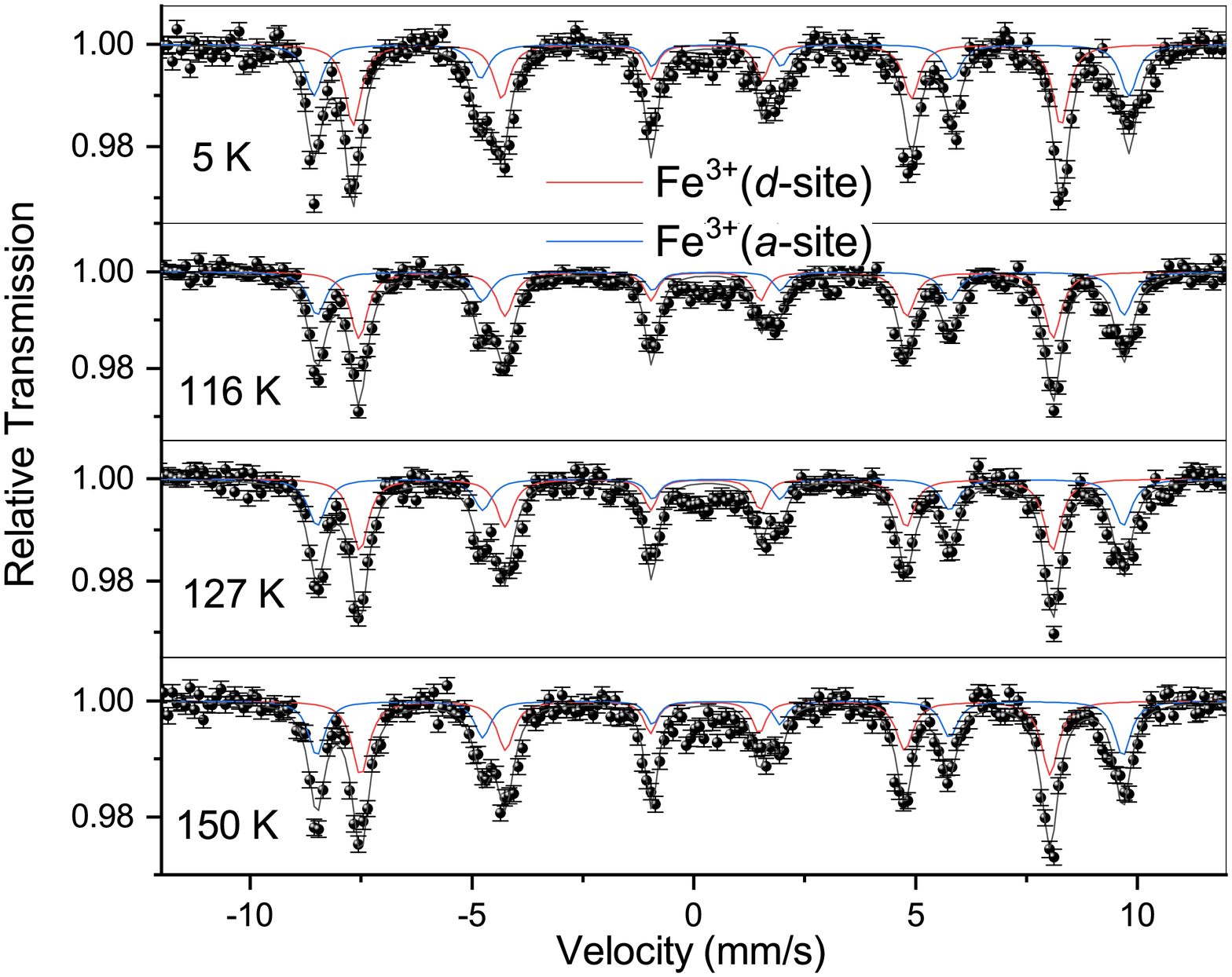}
\caption{$^{57}Fe$ M$\ddot{o}$ssbauer data of Y\textsubscript{1.5}Gd\textsubscript{1.5}Fe\textsubscript{5}O\textsubscript{12} at the selected temperatures. Black symbols represent the experimental data and the black solid line is the best fit to the data.}
\label{fig:MossLT}
\end{figure}

\begin{figure}[b]
\centering
\includegraphics[width=9 cm]{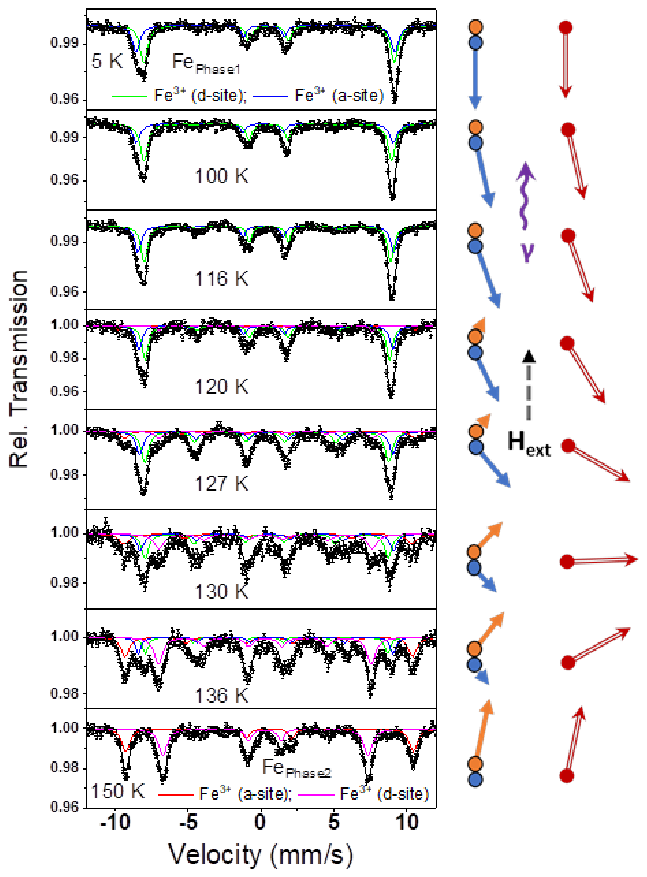}
\caption{In-field $^{57}Fe$ M$\ddot{o}$ssbauer data of Y\textsubscript{1.5}Gd\textsubscript{1.5}Fe\textsubscript{5}O\textsubscript{12} at the selected temperatures under 5 T magnetic field applied parallel to the $\gamma$-rays. Black symbols represent the experimental data and the black solid line is the best fit to the data. Blue arrow represents Fe\textsubscript{phase1} aligned anti-parallel to H\textsubscript{ext}, orange arrow represents Fe\textsubscript{phase2} aligned parallel to H\textsubscript{ext} and the red arrow represents the resultant of these two. H\textsubscript{ext} is the applied external field which is parallel to the incident $\gamma$-rays.}
\label{fig:MossLTHM}
\end{figure} 

\begin{table}[htbp]
\caption{\label{arttype} Hyperfine parameters obtained from the analysis of zero field $^{57}Fe$ M$\ddot{o}$ssbauer spectroscopy data as shown in Figure.~\ref{fig:MossLT}. $\delta$ is centre shift, $\Delta$E is quadrupole shift, H\textsubscript{int} is the internal hyperfine field, A\textsubscript{23} is the area ratio of second and third lines of the respective sextet.}
\begin{ruledtabular}
\begin{tabular}{ccccccc}
T (K)   &$\delta$(mm/s)	   &$\Delta$E(mm/s) & H\textsubscript{int}(T)  & A\textsubscript{23} & \% Area 	&Fe\textsuperscript{3+}  \\	
				&$\pm$0.01					&$\pm$0.02				&$\pm$0.1 							   & 				   				&  $\pm$2       &Site 	  \\
\hline
5 	 		&0.29										&0.01					&49.5                   &2.00        &58   &$\textit{d}$  \\
				&0.57										&0.11					&56.9                   &2.00        &42   &$\textit{a}$ \\	
116	 		&0.27 									&0.01					&48.5                   &2.00        &58   &$\textit{d}$  \\
				&0.56 									&0.10					&56.4                   &2.00        &42   &$\textit{a}$ \\
125	    &0.26 									&-0.01				&48.3                   &2.00        &57   &$\textit{d}$  \\
				&0.56 									&0.13					&56.4                   &2.00        &43   &$\textit{a}$ \\	
127	    &0.26										&0.02					&48.4                   &2.00        &57   &$\textit{d}$  \\
				&0.55										&0.08					&56.3                   &2.00        &43   &$\textit{a}$ \\
136	    &0.26										&0.01					&48.4                   &2.00        &58   &$\textit{d}$  \\
				&0.54										&0.09					&56.3                   &2.00        &42   &$\textit{a}$ \\				
150	 		&0.25										&0.01					&48.2                   &2.00        &59   &$\textit{d}$  \\
				&0.55										&0.08					&56.3                   &2.00        &41   &$\textit{a}$ \\		
								
\end{tabular}
\end{ruledtabular}
\label{tab:parameters1}
\end{table}

\begin{table}[htbp]
\caption{\label{arttype} Representative hyperfine parameters obtained from the analysis of in-field $^{57}Fe$ M$\ddot{o}$ssbauer spectroscopy data as shown in Figure.~\ref{fig:MossLTHM}. H\textsubscript{eff} is the effective field which is the vector sum of H\textsubscript{ext}, H\textsubscript{int} and demagnetizing fields \cite{LTHMMoss_DM}. A\textsubscript{23} is the area ratio of second and third lines of the respective sextet. Fe\textsubscript{phase1} denotes the phase aligned anti-parallel to H\textsubscript{ext} and Fe\textsubscript{phase2} denotes the phase aligned parallel to H\textsubscript{ext} as shown in Figure.~\ref{fig:MossLTHM}.}
\begin{ruledtabular}
\begin{tabular}{ccccccc}
T (K)   & H\textsubscript{eff}(T)  & A\textsubscript{23} & \% Area 	&Fe\textsuperscript{3+} & Phase \\	
				&$\pm$0.1 			& $\pm$0.1	   &  $\pm$2       & Site	&  & \\
\hline
5 	 		&53.2                   &0.00       &58   &$\textit{d}$  & Fe\textsubscript{Phase1}\\
				&54.9                   &0.00       &42   &$\textit{a}$ & Fe\textsubscript{Phase1}\\	
130	 		&51.6                   &1.22       &25   &$\textit{d}$  & Fe\textsubscript{Phase1}\\
				&53.8                   &1.22       &17   &$\textit{a}$ & Fe\textsubscript{Phase1}\\
		    &45.3                   &1.22       &34   &$\textit{d}$  & Fe\textsubscript{Phase2}\\
				&61.1                   &1.22       &24   &$\textit{a}$ & Fe\textsubscript{Phase2}\\							
150	 		&43.6                   &0.10       &59   &$\textit{d}$  & Fe\textsubscript{Phase2}\\
				&61.1                   &0.10       &41   &$\textit{a}$ & Fe\textsubscript{Phase2}\\		
								
\end{tabular}
\end{ruledtabular}
\label{tab:parameters2}
\end{table}

\subsection{$^{57}Fe$ M$\ddot{o}$ssbauer spectroscopy results}

Figure.~\ref{fig:MossLT} shows the temperature dependent $^{57}Fe$ M$\ddot{o}$ssbauer spectra of YGdIG measured across T\textsubscript{Comp}. One can clearly see two components corresponding to Fe\textsuperscript{3+} sublattices located at $\textit{d}$- and $\textit{a}$- sites and the obtained hyperfine parameters match with literature values of garnet \cite{SubstituteYIG_Moss,HFI_Seidel,chinnasamy_Moss}.  Values of H\textsubscript{int}, area fraction and A\textsubscript{23} are shown in Table-1. 
Further, the area fraction of Fe\textsuperscript{3+} at $\textit{d}$- and $\textit{a}$- sites is found to remain same at all the temperatures and the values of A\textsubscript{23} (area ratio of second and third lines in a given sextet) is fixed as 2.0 in accordance with the random distribution of $\theta$ corresponding to the powder samples measured in zero-field conditions \cite{LTHMMoss1}. However, the in-field M$\ddot{o}$ssbauer measurements across T\textsubscript{Comp} with 5 T external magnetic field applied parallel to the $\gamma$-rays, as shown in Figure.~\ref{fig:MossLTHM}, reveal a very interesting information regarding the reversal of Fe\textsuperscript{3+} moments, spin-canting etc., as elaborated in section-IV.

\begin{figure*}[htbp]
\centering
\includegraphics[height= 8cm, width=\textwidth, keepaspectratio]{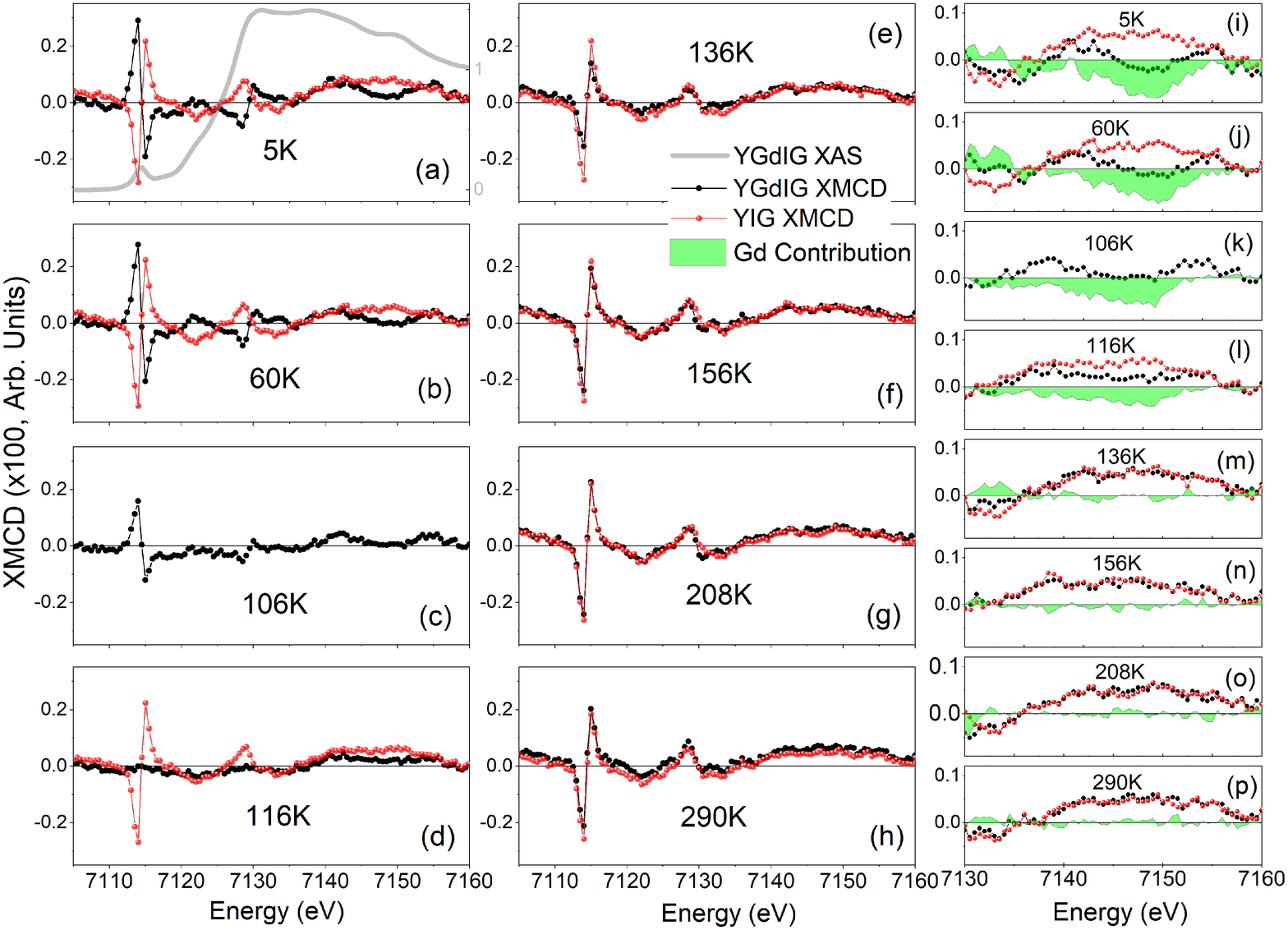}
\caption{(a)-(h) Temperature variation of Fe K-edge X-ray magnetic circular dichroism data of Y\textsubscript{1.5}Gd\textsubscript{1.5}Fe\textsubscript{5}O\textsubscript{12} (YGdIG) and Y\textsubscript{3}Fe\textsubscript{5}O\textsubscript{12} (YIG) at the indicated temperatures. (i)-(p) shows the enlarged view of region of interest depicting the development of Gd contribution at low temperatures. The Gd contribution is extracted by subtracting the YGdIG data from YIG data and its temperature variation is shown in Figure.~\ref{fig:Results2}}
\label{fig:FeXMCD}
\end{figure*} 

\begin{figure}[htbp]
\centering
\includegraphics[width=8 cm]{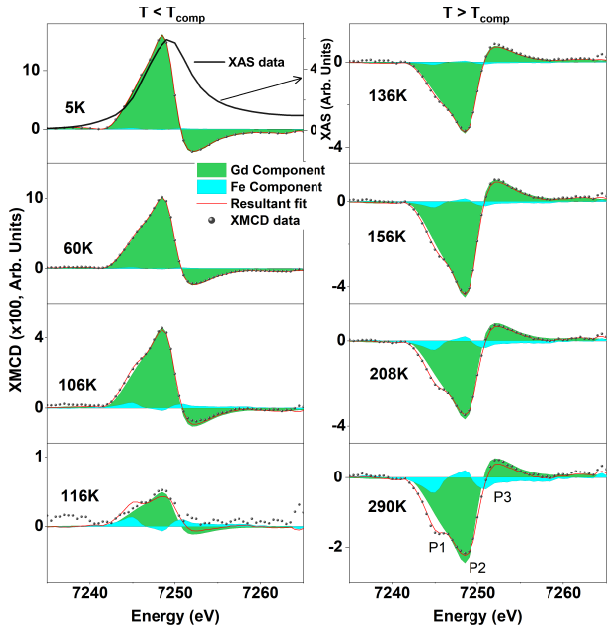}
\caption{Temperature variation of Gd L\textsubscript{3}-edge X-ray magnetic circular dichroism data of Y\textsubscript{1.5}Gd\textsubscript{1.5}Fe\textsubscript{5}O\textsubscript{12}. Symbols represent the experimental data, and red lines are reconstructed spectra using the first two dominating contributions which are shown by shaded areas. Deep green represents Gd- contribution and cyan associated with the Fe- contribution. Labeling of the peaks and the data fitting carried out using singular value decomposition (SVD) method are discussed in the text. X-ray absorption spectra of the sample measured at 5 K is also shown as a representative data.}
\label{fig:GdXMCD}
\end{figure}   

\subsection{Hard x-ray magnetic circular dichroism (XMCD) results}

Figures.~\ref{fig:FeXMCD} and ~\ref{fig:GdXMCD} show the XMCD data measured at the Fe K- and Gd L\textsubscript{3}- edges of YGdIG sample at different temperatures. For the comparison purpose, the Fe K-edge XMCD data of YIG sample is also shown in Figure.~\ref{fig:FeXMCD}. As a representative, XAS data collected at the Fe K-edge and at the Gd L\textsubscript{3}- edges for the YGdIG sample are also shown in Figure.~\ref{fig:FeXMCD} (a) and ~\ref{fig:GdXMCD} (a), respectively. A clear reversal of XMCD signals at Fe K- and Gd L\textsubscript{3}- edges are observed in YGdIG sample when measured at 5 and 290 K i.e., far below and above the T\textsubscript{Comp}. Additional features that are only observed in the Fe K-edge spectra of YGdIG in between 7141 and 7154 eV photon energy, could be due to induced signal of Gd magnetic moments. It may be noted that such a clear signal of rare-earth contribution in Fe K-edge XMCD data was not detected previously in RIG systems, even though the opposite case i.e., Fe contribution in rare-earth L-edge XMCD is observed in many RIG systems \cite{RIG_hardXMCD1, RIG_hardXMCD2, RIG_hardXMCD3, Cornellius_PRL1}. However, unlike RIG systems, the rare-earth contribution was observed at Fe K-edge XMCD data in rare-earth transition metal intermetallics (RTI), which could be due to the fact that the R(5d) electronic orbitals hybridize with the outermost states of absorbing Fe(3d) in RTI systems whereas it is mediated via oxygen in case of RIG systems \cite{Chaboy_PRB2007, Chaboy_PRB2008, Cornellius_PRB2012}. The direct subtraction method from a reference spectrum (YIG, where Y nonmagnetic at $\textit{c}$- site) is employed to separate out the Gd contribution at every temperature, as shown in Figure.~\ref{fig:FeXMCD}(i)-(p), and is further discussed in the following sections.  

Gd L\textsubscript{3}- edge XMCD spectrum can broadly be characterized by the structures labeled as P1, P2, and P3 (Figure.~\ref{fig:GdXMCD}). The peak P1, in the lower energy region that is eventually buried in the profile of peak P2 at low temperatures, can be ascribed to the Fe contribution that becomes visible at temperatures higher than T\textsubscript{Comp} where Gd magnetic moments are weak. Whereas, P2 and P3 mainly originate from the Gd electronic states \cite{RIG_hardXMCD1,RIG_hardXMCD2,RIG_hardXMCD3}. The Gd L\textsubscript{3}- edge XMCD spectra are analyzed with singular value decomposition (SVD) method to extract the Fe\textsuperscript{3+} contribution as discussed in the following section.  

\section{Discussions}

The in-field $^{57}Fe$ M$\ddot{o}$ssbauer data (Figure.~\ref{fig:MossLTHM}) is analyzed considering two Fe sites at temperatures well below and above T\textsubscript{Comp} (5, 100 and 150 K) which correspond to $\textit{d}$- and $\textit{a}$- sites of Fe\textsuperscript{3+}. Representative H\textsubscript{eff} values for some of the temperatures are shown in Table-II. 
.H\textsubscript{ext} will be added (subtracted) to the H\textsubscript{int} of $\textit{a}$- site above (below) the T\textsubscript{Comp} and it will be vice versa for the $\textit{d}$- site as shown schematically in Figure.~\ref{fig:cartoon}. Hence, the H\textsubscript{ext} for the two spectral components corresponding to $\textit{d}$- and $\textit{a}$- sites will be very close to each other below T\textsubscript{Comp}, whereas significant difference will be observed above T\textsubscript{Comp}. As a result of this, one would observe well resolved two sextets corresponding to $\textit{d}$- and $\textit{a}$- sites above T\textsubscript{Comp} and a single broad sextet due to the overlapping components below T\textsubscript{Comp} as shown in Figure.~\ref{fig:MossLTHM}, specifically at temperatures of 150 K and 5 K. Therefore, the obtained H\textsubscript{eff} values which is the vector sum of H\textsubscript{int} and H\textsubscript{ext} unambiguously demonstrates the reversal of Fe\textsuperscript{3+} sublattice across T\textsubscript{Comp} in YGdIG, which can be generalized to all RIG systems exhibiting magnetic compensation.

\begin{figure}[b]
\centering
\includegraphics[height = 15 cm, width = 8 cm, keepaspectratio]{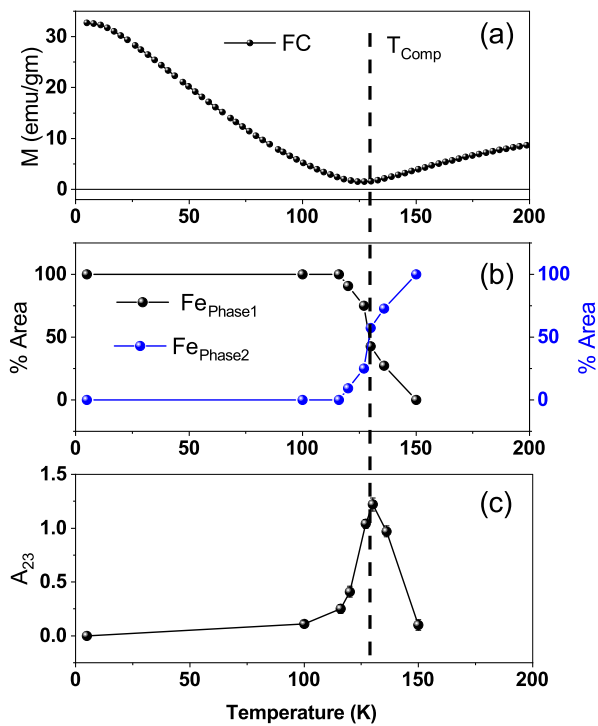}
\caption{(a) M-T data, reproduced from Figure.~\ref{fig:XRD_Mag}(b). Temperature variation of (b) Phase fraction (c) area ratio of second and third line intensities (A\textsubscript{23}) as obtained parameters from the analysis of in-field $^{57}Fe$ M$\ddot{o}$ssbauer data as shown in Figure.~\ref{fig:MossLTHM}.}
\label{fig:Results1}
\end{figure}

However, an inspection of the in-field M$\ddot{o}$ssbauer data close to T\textsubscript{Comp} reveal the presence of more than two components. This is very clearly seen at 127, 130, 136 K and a careful inspection also reveals the presence of these components at 116 and 120 K as shown in Figure.~\ref{fig:MossLTHM}. The spectra could be resolved into four sites corresponding to four Fe\textsuperscript{3+} sublattices of two ferrimagnetic phases of YGdIG i.e., the phases with resultant Fe\textsuperscript{3+} moment aligned along (Fe\textsubscript{phase1}) and opposite (Fe\textsubscript{phase2}) to H\textsubscript{ext} across T\textsubscript{Comp}. As mentioned in the preceding section, the significant contrast between these two phases in terms of H\textsubscript{eff} enable the quantitative study of their evolution with temperature. In view of this, the data at these temperatures is fitted with four sextets and the obtained phase fractions of these two phases (Fe\textsubscript{phase1}, Fe\textsubscript{phase2}) is plotted in Figure.~\ref{fig:Results1}. As the temperature is changing, one can see that there is a gradual change in the area fraction of these phases indicating a continuous spin reversal across T\textsubscript{Comp}.

In addition to this information, the inspection of in-field M$\ddot{o}$ssbauer data close to T\textsubscript{Comp} reveal the presence of considerable intensity for the second and fifth lines (corresponding to $\Delta$m=0 transitions) of a given sextet unlike the data of 5, 100 and 150 K i.e., well below and above T\textsubscript{Comp}. Quantitatively this is estimated by A\textsubscript{23} parameter for given sextet and the value of A\textsubscript{23} is found to be zero at temperatures 5, 100 and 150 K indicating a collinear magnetic structure of Fe\textsuperscript{3+} \cite{LTHMMoss1}. However, the in-field M$\ddot{o}$ssbauer data close to T\textsubscript{Comp} is fitted keeping A\textsubscript{23} parameter as variable and is constrained to be same for all the sites. The obtained variation of A\textsubscript{23} as a function of temperature is also shown in Figure.~\ref{fig:Results1} and it is interesting to note that the magnetic structure deviates from collinear configuration as one approaches T\textsubscript{Comp}. This is considered to be a signature of spin-canting as shown recently from the magnetic circular dichroism \cite{Cornellius_PRL1} and spin Hall magnetoresistance \cite{CantedGdIG_SMR} experiments.  Considering two-sublattice model (Fe\textsuperscript{3+} and R\textsuperscript{3+} sublattices), simulated magnetic field versus temperature (H-T) phase diagram of compensated RIG systems clearly show the region of collinear, canted ferrimagnetic and aligned phases \cite{Neelferri_largeField, Cornellius_PRB2012, CantedGdIG_SMR, Cornellius_PRL1}. 
The present in-field M$\ddot{o}$ssbauer study indicates the presence of such canted phases across T\textsubscript{Comp} unambiguously. Across this region of temperatures, the two Fe\textsubscript{phase1} and Fe\textsubscript{phase2} phases are mixed up with forming similar but opposite angle at a given temperature with respect to H\textsubscript{ext}. As a consequence, resultant Fe moment exhibits a continuous rotation across this canted region in accordance with the two sublattice model as shown schematically in Fig.~\ref{fig:MossLTHM}. It may be emphasized here that even the XMCD data might not be able to distinguish the mixing of these two components as it measures resultant projection of the Fe magnetic moments along the beam direction. 

\begin{figure}[t]
\centering
\includegraphics[width=8 cm]{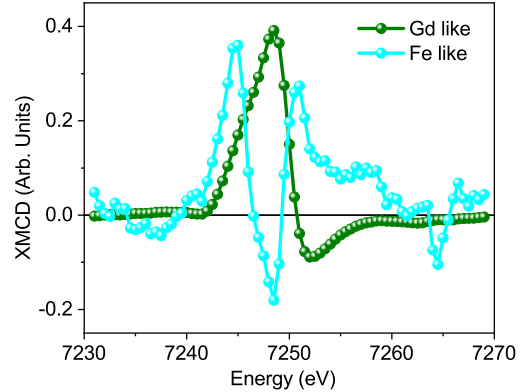}
\caption{Eigen or basis spectra of Gd-like and Fe-like components calculated by singular value decomposition (SVD) method for the Gd L\textsubscript{3}-edge XMCD data set.}
\label{fig:eigenvectors}
\end{figure}

\begin{figure}[htbp]
\centering
\includegraphics[width=8 cm]{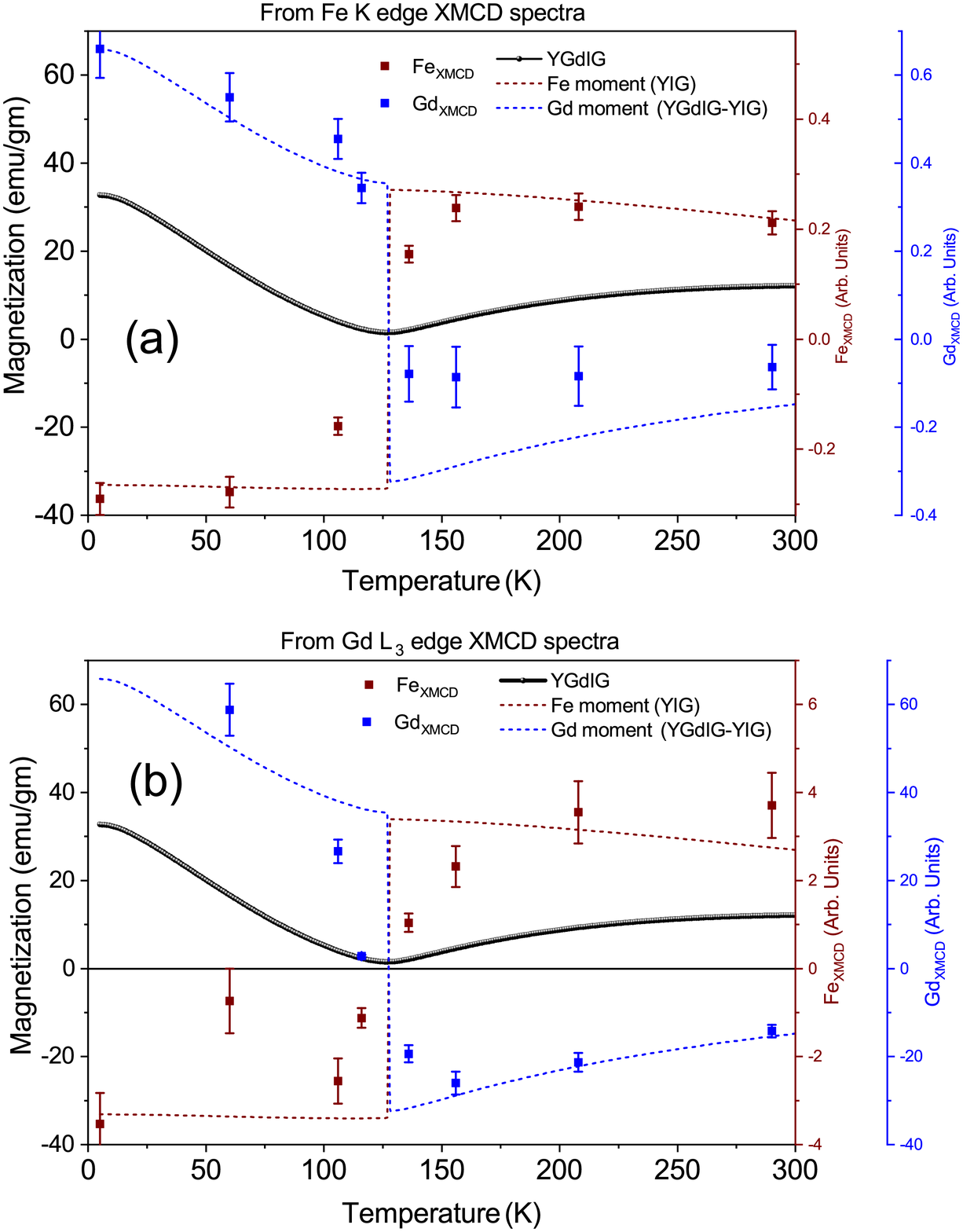}
\caption{The two sublattice (Fe\textsuperscript{3+} and Gd\textsuperscript{3+}) magnetization values (proportional to XMCD amplitude) obtained from the (a) Fe K- and (b) Gd L\textsubscript{3}- edge XMCD data analysis. The values are shown by symbols. Temperature variation of bulk magnetization data of YGdIG sample (Fig.~\ref{fig:XRD_Mag}) and the calculated contribution of Fe\textsuperscript{3+} and Gd\textsuperscript{3+} sublattices to the bulk magnetization are shown in both frames. M-T data of YIG is considered as due to Fe\textsuperscript{3+}, except that the sign of the data is changed accordingly across T\textsubscript{Comp}. }
\label{fig:Results2}
\end{figure} 

Apart from showing a clear reversal of the magnetic signals at the Fe K- and the Gd L\textsubscript{3}- edges in YGdIG acorss the T\textsubscript{Comp}, the XMCD spectra reported in Figure.~\ref{fig:FeXMCD}and ~\ref{fig:GdXMCD} are employed to extract quantitative information about both the sublattices magnetization from a single absorption edge measurement. Mainly two approaches are used in literature to extract the spectral contributions of different sublattices from such complicated XMCD data \cite{Chaboy_PRB2007,RIG_hardXMCD3}. Subtraction method in which the magnetic dichroic signal from the investigated system is subtracted from a reference sample has been extensively exploited and has provided very reliable interpretation \cite{Chaboy_PRB2007}.In our present work, XMCD spectra obtained at the Fe K-edge for the YIG sample are considered as reference to estimate the magnetic contribution from the Gd ions at the Fe K-edge. Using this simple methodology, the Gd magnetic contribution can be obsevred via XMCD at the Fe K-edge as shown in Figure.~\ref{fig:FeXMCD}(i)-(p). 

In order to check the reliability of thus obtained Gd\textsuperscript{3+} contribution, the total area of these features observed at the Fe K post-edge are compared with the temperature variation of Gd\textsuperscript{3+} moment, as calculated from temperature dependent bulk magnetization data of YGdIG and YIG.The results using this approach are summarized in Fig.~\ref{fig:Results2}(a). The bulk magnetization data of YIG is considered to have its origin basically from the Fe\textsuperscript{3+} ions since we can assume that both Y\textsuperscript{3+} and O\textsuperscript{2-} ions have negligible magnetic moments and their hybridization with the Fe\textsuperscript{3+} ions does not play an important role to the total magnetism of the system. Therefore, subtracting bulk magnetization data of YGdIG from the YIG data (Fig. 2), results magnetic signal which is primarily from the Gd\textsuperscript{3+} ions. The sign of the obtained sublattice magnetization as a function of temperature is changed according to their magnetization direction w.r.t H\textsubscript{ext} across T\textsubscript{Comp}(Fig.~\ref{fig:Results2} dashed line). The extracted Gd\textsuperscript{3+} XMCD amplitude (blue dot Fig.~\ref{fig:Results2}(a)) from Fe K edge XMCD data (using direct subtraction method) almost follow the trend of Gd\textsuperscript{3+} moment variation as a function of temperature below the T\textsubscript{Comp}. However, above T\textsubscript{Comp} Gd\textsuperscript{3+} magnetic contribution to Fe K-edge XMCD data is weaker and hence giving discrepancy to the comparison. We have also compared Fe XMCD value (red symbols in Fig.~\ref{fig:Results2}(a), peak height of the XMCD dispersion spectra located at the pre-edge region ($\approx$ 7115 eV) with the temperature variation of the Fe-only magnetization data from the YIG sample and it was observed similar trend (red dashed line) as shown in Fig.~\ref{fig:Results2}(a).

As mentioned above the direct subtraction method or linear combination fit according to individual magnetic moment value as function of temperature are mostly employed in literature to decompose the XMCD spectra of RIG and RTI systesm \cite{Chaboy_PRB2007, RIG_hardXMCD3}. However, recently it is shown that the application of singular value decomposition (SVD) rationalizes previous approaches in a more general framework and simplifies the exploration of magnetic phase diagram of such compounds \cite{Cornellius_PRL1, Cornellius_PRL2}. One can analyze the shape and amplitude of the hidden components in XMCD data from correlated data set. Cornellius et al., successfully analyzed and separated Fe contribution from the Er L\textsubscript{2,3} edges using SVD method \citep{Cornellius_PRL1}. The same procedure is adopted in the present work also to de-convolute Fe\textsuperscript{3+} contribution from temperature dependence Gd L\textsubscript{3} edge XMCD data set (Figure.~\ref{fig:GdXMCD}). 

According to SVD theorem any data matrix $\textbf{A}$(m$\times$n) can be decomposed into three matrices as \textbf{A=U$\Sigma$V\textsuperscript{T}} where ${U}(m\times m)$ and $V(n\times n)$ are orthogonal matrices and $\Sigma(m\times n)$ is diagonal matrix of singular values. We have used eight XMCD spectra of Gd L\textsubscript{3}- edge over the temperature range (5-300 K) to form data matrix $\textbf{A}$ and MATLAB software is used to perform the SVD on the data matrix $\textbf{A}$ to find the $\textbf{U, V, $\Sigma$}$. Only the first two eigenvectors or basis which are dominated over the noise level are used to reconstruct the original data as shown in Figure.~\ref{fig:eigenvectors}. The reconstructed plot along with separated Fe and Gd like components to the original Gd L\textsubscript{3}- edge XMCD data are shown in Figure.~\ref{fig:GdXMCD}. It is to be mentioned that spectral features of Fe contribution from Gd L\textsubscript{3}- edge XMCD obtained by SVD in the present work match with previous literature\cite{Cornellius_PRL1, RIG_hardXMCD3} and the XMCD signal amplitude (red symbols), as shown in Fig.~\ref{fig:Results2}(b), is also following the similar temperature variation as of Fe sublattice magnetization data (red dashed line) from YIG in terms of its magnitude and direction. The SVD separated Gd XMCD amplitudes (blue symbols) at indicated temperature and corresponding temperature variation of Gd only sublattice magnetization (blue dashed line) are  also shown in the Fig.~\ref{fig:Results2}(b).
 
\section{Conclusions}

In conclusion, in the present work in-field $^{57}Fe$ M$\ddot{o}$ssbauer spectroscopy is employed to demonstrate the Fe\textsuperscript{3+} spin reversal and signatures of spin-canting across the magnetic compensation temperature (T\textsubscript{Comp}) in Y\textsubscript{1.5}Gd\textsubscript{1.5}Fe\textsubscript{5}O\textsubscript{12}. The M$\ddot{o}$ssbauer data also clearly demonstrate the continuous rotation of Fe\textsuperscript{3+} moment across the T\textsubscript{comp}, which is nothing but a second order field induced phase transition.  Inversion of sublattice spin across the T\textsubscript{Comp} is also confirmed by Fe K- and Gd L\textsubscript{3}- edge XMCD data.
The quantitative estimation of the two sublattice contribution viz., Fe\textsuperscript{3+} and Gd\textsuperscript{3+} to the net magnetization is separated out from XMCD spectra collected from either Fe K- or Gd L\textsubscript{3}- edge using direct subtraction method from reference spectrum and SVD method, respectively i.e., in RIG systems one can probe both Fe\textsuperscript{3+} and Gd\textsuperscript{3+} magnetism from only single edge XMCD measurements in the hard x-ray region.  
\\

\maketitle \section{Acknowledgments}

We acknowledge DESY (Hamburg, Germany), a member of the Helmholtz Association HGF, for the provision of experimental facilities. Parts of this research were carried out at beamline P09. Beamtime was allocated for proposal I-20180676 within the India@DESY collaboration in Photon Science. VRR and MK would like to acknowledge the financial support by the Department of Science and Technology (Government of India) provided within the framework of the India-DESY Collaboration. MK thank Mr. Anil Gome and Mr. Deepak Prajapat for the help with LTHM $^{57}Fe$ M$\ddot{o}$ssbauer measurements.  Mr. Kranti Kumar Sharma is thanked for M-T data.


\bibliography{Reference_Garnet}

\end{document}